\documentclass[10pt,preprintnumbers, floatfix, letterpaper, onecolumn,aps,prd,epsfig,nofootinbib,nat_bib,longbibliography]{revtex4-1}
\usepackage{graphicx}
\usepackage[utf8]{inputenc}
\usepackage[T1]{fontenc}
\usepackage{latexsym}
\usepackage{braket}
\usepackage{epstopdf}
\usepackage{latexsym}
\usepackage{amssymb}
\usepackage{amsmath}
\usepackage{mathtools}
\usepackage{color}
\usepackage{mathrsfs}
\usepackage{xparse}
\usepackage{enumerate}
\usepackage{multirow,tabularx,boldline,array}
\usepackage{bbding}
\usepackage{enumitem}
\usepackage{pifont}
\usepackage[center]{subfigure}
\usepackage[
            pdfstartview=FitH,
            bookmarksnumbered=true,
            bookmarksopen=true,
            colorlinks,
            linkcolor=blue,
            anchorcolor=green,
            citecolor=blue
            ]{hyperref}

\newcommand{\mpref}[1]{Figure.\ref{#1}}

\begin{document}

\title{Islands in Kerr-Newman Black Holes}
\author{Ming-Hui Yu}
	\email{yuminghui@shu.edu.cn}
	\affiliation{Department of Mathematics, Shanghai University, 99 Shangda Road, Shanghai, 200444, China}
		\author{Xian-Hui Ge}
	\email{gexh@shu.edu.cn; Corresponding author}
   \affiliation{Department of Physics, Shanghai University, 99 Shangda Road, Shanghai, 200444, China}

\date{\today}

\begin{abstract}
We investigate the information paradox in the four-dimensional Kerr-Newman black hole by employing the recently proposed island paradigm. We first consider the quantum field in the four-dimensional Kerr-Newman spacetime. By employing the near-horizon limit, we demonstrate that the field can be effectively described by a reduced two-dimensional field theory. Consequently, the formula of entanglement entropy in CFT$_2$ can be naturally adapted to this reduced two-dimensional theory. Under the framework of this reduced two-dimensional theory, we show that the entanglement entropy of radiation for the non-extremal case satisfies the unitarity in the later stage of the appearance of the entanglement islands. We further examine the impact of angular momentum and charges on the Page time and the scrambling time. Both quantities increases as the angular momentum increases, while decreases as the charge increases. At last, we consider the near extremal case. Resort to the Kerr/CFT correspondence, the near-horizon geometry of near extremal Kerr-Newman black holes can be taken account for  a warped AdS geometry.  In this scenario, the low-energy effective degrees of freedom are dominated by the Schwarzian zero mode, resulting in a one-loop correction to the partition function. The entanglement entropy is subsequently recalculated under the thermodynamic with corrections. Through explicit calculations, we finally find that the Page time and the scrambling time exhibits quantum delays. This strongly suggests that the near extremal geometry is governed by the Schwarzian dynamics, in which quantum fluctuations result in a reduced rate of information leakage. Our findings further substantiate the conservation of information and extend the applicability of the island paradigm to the most general stationary spacetime background.
\end{abstract}

\maketitle
\tableofcontents
\newpage

\section{Introduction} \label{sec1}
 \renewcommand{\theequation}{1.\arabic{equation}}\setcounter{equation}{0}

\par The quantum properties of the black hole have significantly contributed to the development of a consistent theory of quantum gravity. Among these properties, one of the most intriguing is Hawking radiation, which first proposed by Stephen Hawking \cite{hr}. Since the Hawking radiation involves classical general relativity (GR) and quantum field theory, which provides a profound insight into the quantum gravity. However, Hawking radiation will eventually lead black holes to evaporate and disappear. More specifically, a black holes formed from the pure state eventually evolves into a mixed state \cite{mixed}.
This process violates the principle of quantum mechanics known as the unitarity, which leads to information loss \cite{paradox}. The requirement of the unitarity suggests that the final state of a black hole formed by the pure state must also remain in a pure state. Consequently, this has given rise to the well-known black hole information paradox \cite{paradox}.

\par The von Neumann entropy, or the entanglement entropy, is utilized to express the quantification of the amount of information loss. Hawking's calculations in that the entanglement entropy of radiation continuously increases with time and eventually exceeds the Bekenstein-Hawking entropy bound \cite{bound} in the later stages of black hole evaporation. However, the unitary evaporation of black holes is expected to be accompanied by a particular form of the evolution of the entanglement entropy. Specifically, the entanglement entropy must asymptotically decrease to zero in end of evaporation. Based on this principle, Page demonstrates that the so-called Page curve \cite{pc1} must be satisfied during the evaporation process of the entanglement entropy of Hawking radiation. Therefore, the key of resolving the black hole information paradox hinges on successfully reproducing the corresponding Page curve in the specific theoretical framework \cite{pc2}.

\par Calculating the Page curve has been a very challenging task until the proposal of the AdS/CFT duality \cite{adscft}. This theory suggests that the gravitational theory in anti-de Sitter (AdS) spacetime can be equivalently described by the conformal field theory (CFT) on its boundary. This finding offers conclusive evidence that the evaporation process of black holes in AdS spacetime is unitary. Recently, this field has a significant breakthrough \cite{bulk entropy,entanglement wedge,island rule,review}. The related studies indicates that the Page curve can be calculated by the following island formula based on the quantum extremal surface (QES) prescription \cite{qes}:
\begin{equation}
\begin{split}
S_R &= \text{min} \Big[ \text{ext} \big (S_{\text{gen}} \big) \Big]  \\
    &= \text{min} \bigg[ \text{ext} \bigg( \frac{\text{Area}(\partial I)}{4G_N} +S_{\text{CFT}} (R \cup I) \bigg) \bigg], \label{island formula}
\end{split}
\end{equation}
 where $S_{\text{gen}}$ is denoted as the generalized entropy. It can simply be regarded as an extension of the Ryu-Takayanagi formula \cite{rt,hrt,qrt}. The generalized entropy is composed of the sum of the area term and the entropy associated with the quantum matter field. Here $R$ represents the radiation region, $I$ represents the island region with the boundary $\partial I$. The key to the island formula \eqref{island formula} lies in that one first extremizes the generalized entropy (ext) to obtain the position of QES, and then select the minimum value among all the candidates as the entanglement entropy of the radiation. This is also known the island paradigm \cite{review}. Furthermore, the island formula \eqref{island formula} can be rigorously derived from the Euclidean path integral through the application of the replica trick \cite{replica1,replica2,replica3}. Detailed calculations reveal that the emergence of the entanglement islands is associated with the replica wormholes saddle that dominates the evaporation at late times.

\par At present, the island paradigm has not only be applied to the initial evaporative Jackiw-Teitelboim gravity \cite{jtgravity1,jtgravity2} but also extends to scenarios involving eternal black holes \cite{eternalbh} and various specific spacetime background \cite{jtisland,2d1,2d2,2d3,2d4,2d5,2d6,ymh2d,ymhjt,2d7,ds1,ds2,ds3,ds4,ds5,cosmology1,cosmology2,cosmology3,cosmology4,cosmology5,cosmology6,cosmology7,wedgeholo1,wedgeholo2,wedgeholo3,wedgeholo4,wedgeholo5,wedgeholo6,negative,reflectedentropy1,reflectedentropy2,other1,other2,other3,other4,other5,other6,sch,rn,kerr,kerrds,extremalbh1,extremalbh2,other7,other8,other9,other10,other11,other12,other13,other14,other15,other16,other17,other18,other19,ymhghs,ymhbtz,ymhacc,ymhrn,ymhqfc,ymhmodular,other20,other21,other22,other23,other24,other25,other26,other27,other28,other29,other30,other31,other32,other33,other34,other35,other36,other37,other38,other39,other40,other41,other42,other43,other44,other45,other46,other47,other48}. However, in these studies, nearly all have focused on the case of black holes that are \emph{static spherically symmetric} \cite{jtisland,2d1,2d2,2d3,2d4,2d5,2d6,ymh2d,ymhjt,2d7,other1,other2,other3,other4,other5,other6,sch,rn,kerr,kerrds,extremalbh1,extremalbh2,other7,other8,other9,other10,other11,other12,other13,other14,other15,other16,other17,other18,other19,ymhghs,ymhbtz,ymhacc,ymhrn,ymhqfc,ymhmodular,other20,other21,other22,other23,other24,other25,other26,other27,other28,other29,other30,other31,other32,other33,other34,other35,other36,other37,other38,other39,other40}. There are few reports on \emph{non-spherically symmetric} stationary black holes \cite{ymhbtz,kerr,kerrds}. For the most general type in four-dimensional spacetime, known as the Kerr-Newman black hole. Its metric describes a rotating, charged mass and represents the most general solution to the Einstein's equations in GR. Therefore, it holds significant theoretical importance in the mathematical framework of GR and extends beyond. Correspondingly, resolving the information paradox in this black hole is a critical issue. This study can also offer valuable insights into other four-dimensional scenarios.

\par Since the analytical expression of the entanglement entropy is difficult to obtain for higher-dimensional spacetime, one usually adopts the s-wave approximation to calculate the entanglement entropy of the two-dimensional matter field \cite{sch}. However, for four-dimensional Kerr-Newman black holes, the scalar field in this metric can be reduced to an effective two-dimensional theory at the near-horizon limit \cite{reduction}. Consequently, the s-wave approximation retains its validity under this case. Furthermore, we neglect the back-reaction of Hawking radiation on spacetime in order to maintain the stability of the spacetime structure. On the other hand, due to the presence of the angular momentum $J$ and the charge $Q$, black holes emits the non-thermal superradiacne \cite{suprad}. At this stage, the whole evaporation process becomes quite complicated. For the sake of simplicity, we take into account the large mass limit: $M \gg L \sim Q$. Namely, the Hawking radiation dominates in the evaporation, rendering the superradiance is negligible.

\par The structure of this paper is organized as follows. In section \ref{sec2}, we briefly review the fundamental properties of the Kerr-Newman black hole and introduce the standard technique that can transform a quantum field in the four-dimensional Kerr-Newman spacetime into an effective two-dimensional description. This approach establishes an effective two-dimensional metric near the event horizon, which facilitates the subsequent calculation of the entanglement entropy. In section \ref{sec3}, we rigorously calculate the entanglement entropy of the radiation emitted by non-extremal Kerr-Newman black holes and derive the corresponding Page curve. Furthermore, we analyze the effects of the charge and the angular momentum on both the Page time and the scrambling time. In section \ref{sec4}, we focus on the near extremal cases. Based on the Kerr/CFT correspondence, the near extremal Kerr-Newman geometry approximates to a warped AdS$_3$ at the near-horizon region. By incorporating the one-loop correction to the modified thermodynamics through the zero mode associated with Schwarzian dynamics, we arrive at a final result of Page time and scrambling time. We find these quantities will be significantly delayed due to the one-loop correction and provide the correct result in the near extremal limit. Finally, the section \ref{sec5} summarizes the conclusions and provides further discussions.

\section{Quantum Fields in Kerr-Newman Spacetime}  \label{sec2}
\renewcommand{\theequation}{2.\arabic{equation}}\setcounter{equation}{0}

\par In this section, we provide a concise review of the Kerr-Newman spacetime and its distinctive properties. Subsequently, we provide a detailed description of the method used to reduce a four-dimensional theory to an effective two-dimensional theory by employing the near-horizon limit. Then, the subsequent calculation of entanglement entropy can be simplified.

\subsection{Kerr-Newman Black Black Holes}

\par  In this subsection, we provide some useful relations of Kerr-Newman black holes, which are used in the following content. The metric for the rotating charged Kerr-Newman spacetime in the Boyer-Lindquist coordinate are written as follows
\begin{equation}
ds^2 = - \frac{\Delta-a^2 \sin^2 \theta}{\Sigma} dt^2 - \frac{2a \sin^2 \theta (r^2+a^2-\Delta)}{\Sigma} dt d\phi + \frac{(r^2+a^2)^2 - \Delta a^2 \sin^2 \theta}{\Sigma} \sin^2 \theta d\phi^2 + \frac{\Sigma}{\Delta} dr^2 +\Sigma d\theta^2, \label{kn metric1}
\end{equation}
where these notations are respectively defined by
\begin{subequations}
\begin{align}
a              &\equiv \frac{L}{M},  \label{parameter1} \\
\Sigma         &\equiv r^2 + a^2 \cos^2 \theta, \label{parameter2} \\
\Delta         &\equiv r^2 -2Mr +a^2 +Q^2 = (r-r_+)(r-r_-).  \label{parameter3}
\end{align}
\end{subequations}
In the above equations, $M$, $J$, $Q$ and $r_{\pm}$ correspond to mass, angular momentum, charge and event(outer)/inner horizons, respectively. Moreover,
\begin{equation}
r_{\pm} = M \pm \sqrt{M^2-a^2-Q^2}.  \label{horizons}
\end{equation}
\par For the general case, the condition $0 \le a^2+Q^2 <M^2$ describes the non-extremal black hole. However, it should to notice that two horizons coincide when $M^2=a^2+Q^2$. It represents for the extremal case. The surface gravity $\kappa$ is related to the Killing vector $\xi ^{\mu}$ of the event horizon
\begin{equation}
\kappa(r) = \sqrt{- \frac{1}{2} \nabla^{\mu} \xi^{\nu}  \nabla_{\mu} \xi_{\nu} } = \frac{r_+-r_-}{2(r^2+a^2)}. \label{surface gravity}
\end{equation}
Here $\xi ^{\mu}= (\partial_t)^{\mu} + \Omega_H (\partial_{\phi})^{\mu}$, where $\Omega_H=\frac{a}{r_+^2 +a^2}$ is the angular velocity at the event horizon. The symbol $\nabla_{\mu}$ is denoted as the covariant derivative operator. Then the Hawking temperature is derived by
\begin{equation}
T_H =\frac{\kappa(r_+)}{2\pi} = \frac{r_+-r_-}{4\pi (r_+^2+a^2)}.  \label{temperature}
\end{equation}
The area of the event horizon is given by
\begin{equation}
\begin{split}
\mathcal{A} &= \int \sqrt{-g} d\theta d\varphi  =\int_{0}^{2\pi} d\varphi \int_0^{\pi} d \theta (r^2+a^2) \sin \theta \\
            &=4\pi (r_+^2 +a^2).  \label{area}
\end{split}
\end{equation}
Thus, the Bekenstein-Hawking entropy is read off as
\begin{equation}
S_{\text{BH}} \equiv \frac{\mathcal{A}}{4G_N} =\frac{\pi (r_+^2+a^2)}{G_N}.  \label{bh entropy}
\end{equation}
It needs to be emphasized that, in the extremal case, both the surface gravity \eqref{surface gravity} and the Hawking temperature \eqref{temperature} are vanishing. We will discuss the extremal case in detail in the subsequent section.

\subsection{The Effective Two-Dimensional Theory}

\par In this subsection, we demonstrate that the quantum filed in four-dimensional Kerr-Newman spacetime can be transformed to a two-dimensional theory by imposing the near-horizon limit.

\par For convenience, consider a complex scalar field $\phi^{\star}$ in Kerr-Newman spacetime. This action is written as
\begin{equation}
I= \int d^4x \sqrt{-g} g^{\mu \nu} (\partial_{\mu} +i e V_{\mu} ) \phi^{\star} (\partial_{\nu} - i e V_{\nu}) \phi + I_{\text{int}}.  \label{action1}
\end{equation}
where the first term is denoted as the kinetic term, the second term is the interaction terms. The gauge field $V_{\mu}$ is given by $\Big(-\frac{Qr}{r^2+a^2}, 0, 0, 0 \Big)$. Then, we substitute the Kerr-Newman metric \eqref{kn metric1} to the action \eqref{action1}, yielding to \cite{reduction}
\begin{equation}
\begin{split}
I &= \int dt dr d\theta d\varphi  \sin \theta  \phi^{\star} \bigg [ \bigg( \frac{(r^2+a^2)}{\Delta} -a^2 \sin^2 \theta \bigg) \bigg( \partial_t +\frac{ieQr}{r^2+a^2} \bigg)^2 +2ia \bigg( \frac{r^2+a^2}{\Delta} -1 \bigg) \bigg( \partial_t + \frac{ieQr}{r^2+a^2}   \bigg) \hat{L}_z  \\
&- \partial_r \Delta \partial_r +\hat{L}^2  - \frac{a^2}{\Delta} \hat{L}_z^2  \bigg] \phi + I_{\text{int}},  \label{action2}
\end{split}
\end{equation}
where we have used the angular momentum operator
\begin{equation}
\hat{L} = - \frac{1}{\sin \theta} \partial_{\theta}  \sin \theta \partial_{\theta}  -\frac{1}{\sin^2 \theta} \partial_{\varphi}^2,  \qquad \hat{L}_z = -i \partial_{\varphi}.  \label{operator}
\end{equation}
Then we expand the scalar field $\phi$ in terms of the spherical harmonics: $\phi = \sum_{l,m} \phi_{lm} (t,r) Y_{lm}(\theta,\varphi)$, we obtain
\begin{equation}
\begin{split}
I &= \int dt dr d\theta d\varphi  \sin \theta   \sum_{l^{\prime}, m^{\prime} } \phi^{\star}_{l^{\prime} m^{\prime}} Y^{\star}_{l^{\prime} m^{\prime}} \Bigg[  \frac{(r^2+a^2)^2}{\Delta} \bigg( \partial_t +\frac{ieQr}{r^2+a^2}  \bigg)^2 -a^2 \sin^2 \theta \bigg( \partial_t +\frac{ieQr}{r^2+a^2}  \bigg)^2  +2ima \frac{r^2+a^2}{\Delta}   \\
&\times \bigg( \partial_t + \frac{ieQr}{r^2+a^2}  \bigg)  -2ima \bigg( \partial_t + \frac{ieQr}{r^2+a^2} \bigg) -\partial_r \Delta \partial_r  +l(l+1) - \frac{m^2a^2}{ \Delta}  \Bigg]  \sum_{l,m} \phi_{lm} Y_{lm} + I_{\text{int}}. \label{action3}
\end{split}
\end{equation}
In the above equation, the eigenvalue equation for $\hat{L}^2$ and $\hat{L}_z$ is used to simplify
\begin{equation}
\hat{L}^2 Y_{lm} = l(l+1) Y_{lm},  \qquad   \hat{L}_z Y_{lm} = m Y_{lm}, \label{eigenvalue}
\end{equation}
with $l$ and $m$ represents the azimuthal quantum number and the magnetic quantum number, respectively. Now we define the tortoise coordinate as a service for subsequent calculation
\begin{equation}
\begin{split}
r_{\star}(r) &\equiv  \int \frac{dr}{f(r)} = \int \frac{r^2+a^2}{\Delta} dr \\
             & = r+  \frac{(a^2+r_+^2) \log |r-r_+| - (a^2+r_-^2)  \log |r-r_-|}{ r_+-r_-}.  \label{tortoise1}
\end{split}
\end{equation}
Furthermore, we employ the near-horizon limit for the action \eqref{action3}. Near the event horizon $r \to r_+$, we have $f(r \simeq r_+) =0$ and only the dominant term in \eqref{action3} is left, which yields
\begin{equation}
\begin{split}
I(r_{\star}) &= \int dt dr_{\star} d\theta d\varphi  \sin \theta  \sum_{l^{\prime} m^{\prime}}  \phi^{\star}_{l^{\prime} m^{\prime}}  Y_{l^{\prime} m^{\prime}}^{\star}   \Bigg[ (r^2+a^2)  \bigg( \partial_t  + \frac{ieQr}{r^2+a^2}  \bigg)^2 +2ima \bigg( \partial_t  + \frac{ieQr}{r^2+a^2}  \bigg)   \\
& - \partial_{r_{\star}} (r^2+a^2) \partial_{r_{\star}} - \frac{m^2a^2}{r^2+a^2}  \Bigg]  \sum_{l^{\prime} m^{\prime}} \phi_{lm}  Y_{lm}.  \label{action4}
\end{split}
\end{equation}
The interaction term $I_{\text{int}}$ is discarded due to the fact that the kinetic term is dominate and the effective theory becomes to a high-energy state near the horizon. Then, we use the orthogonal condition for the spherical harmonics: $\int d \theta d \varphi  \sin \theta  Y_{l^{\prime} m^{\prime}}^{\star} Y_{lm}  = \delta_{l^{\prime}l}  \delta_{m^{\prime}m}$ in the expression \eqref{action4} and rewrite it in terms of $r$
\begin{equation}
I(r) = - \sum_{l,m}  \int dt dr (r^2+a^2) \phi^{\star}_{lm} \bigg[  -\frac{r^2+a^2}{\Delta} \bigg( \partial_t + \frac{ieQr}{r^2+a^2}  +\frac{ima}{r^2+a^2} \bigg)^2  + \partial_r  \frac{\Delta}{r^2+a^2} \partial_r \bigg] \phi_{lm}.  \label{action5}
\end{equation}
Eventually, we can regard $\phi_{lm}$ as a two-dimensional complex scalar field in a two-dimensional spherical symmetric metric $g_{\mu \nu}$ with the dilaton $\Phi$ and two $U(1)$ gauge fields $\text{U}$, $\text{V}$:
\begin{subequations}
\begin{align}
\phi &= r^2+a^2,  \\
-g_{tt} &= \frac{1}{g_rr} =f(r), \qquad g_{rt}=0,   \\
\text{U}_t &= - \frac{a}{r^2+a^2},  \qquad  \ \  \text{U}_r=0,  \\
\text{V}_t &= - \frac{Qr}{r^2+a^2},  \qquad \ \ \text{V}_r=0.
\end{align}
\end{subequations}
In fact, the gauge filed $\text{V}_{\mu}$ is the original gauge field in the action \eqref{action1}. The other $\text{U}_{\mu}$ is the induced gauge field related to the isometry along the $\varphi$ direction with the $U(1)$ charge $m$. Therefore, the gauge potential $A_t$ is the sum of them
\begin{equation}
A_t =e \text{V}_t  + m \text{U}_t  = - \bigg( \frac{eQr+ma}{r^2+a^2}  \bigg),  \qquad A_r=0.   \label{potential}
\end{equation}

\par In the end, through this set of procedures, we utilize the two-dimensional spherically symmetric effective theory at the near-horizon limit $(r \sim r_+)$ to describe the behavior of quantum fields in four-dimensional non-spherical symmetric Kerr-Newman spacetime \eqref{kn metric1} :
\begin{equation}
\begin{split}
ds^2_{\text{eff}} &= -f(r) d\tau^2 +f^{-1}(r)dr^2  \\
&=- \frac{(r-r_+)(r-r_-)}{r^2+a^2}  d\tau^2  + \frac{r^2+a^2}{(r-r_+)(r-r_-)} dr^2.  \label{2d metric1}
\end{split}
\end{equation}
The correctness of this result can be verified by calculating the Hawking temperature, which is given by
\begin{equation}
T_H \equiv \frac{f^{\prime}(r)}{4\pi} \bigg|_{r=r_+} =\frac{r_+-r_-}{4\pi (r_+^2 +a^2)}.  \label{temperature2}
\end{equation}
This temperature derived by the metric \eqref{2d metric1} is consistent with the temperature \eqref{temperature}. Therefore, we use the effective theory \eqref{2d metric1} to calculate the entanglement entropy in the following content.

\subsection{Conformal Flat Form for non-extremal Case}

\par
\begin{figure}[htb]
\centering
\includegraphics[scale=1.2]{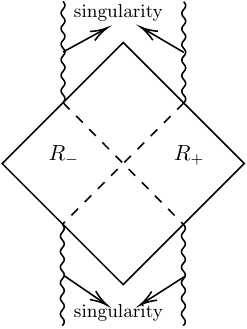}
\caption{The Penrose diagram for Kerr-Newman black holes. These dash lines are labeled by event horizons, which divides the whole spacetime into four wedges. $R_{\pm}$ represents the left and right wedge where Hawking radiation exists.}
\label{penrose1}
\end{figure}

 Now, we have the effective two-dimensional metric \eqref{2d metric1} to describe the quantum filed in four-dimensional Kerr-Newman spacetime. In order to facilitate subsequent calculations and obtain the extension of spacetime, we need to employ the Kruskal transformation. In this section, we only focus on the non-extremal black hole. For the extremal case, we will discuss this in the section \ref{sec4}. For the Kerr-Newman spacetime, the corresponding Penrose diagram is shown in \mpref{penrose1}.

\par For the non-extremal case, the tortoise coordinate is defined by \eqref{tortoise1}. The, define the null coordinate $\{u,v \}$: $u = \tau -r_{\star}$, $v=\tau+r_{\star}$. Accordingly, the Kruskal coordinate $\{U, V\}$ that can eliminate the coordinate singularities are written as
\begin{equation}
\begin{split}
\text{Left Wedge}:  \ \ U\equiv +e^{-\kappa u};  \qquad V\equiv - e^{+\kappa v}.  \\
\text{Right Wedge}: \ \ U\equiv -e^{-\kappa u};  \qquad V\equiv + e^{+\kappa v}.  \label{kruskal}
\end{split}
\end{equation}
Under this transformation, the effective metric \eqref{2d metric1} becomes to a conformal flat form
\begin{equation}
ds^2 = - \frac{dU dV}{\Omega^2(r)},  \label{conformal flat}
\end{equation}
with the conformal factor $\Omega(r)$
\begin{equation}
\Omega(r) \equiv \frac{\kappa e^{\kappa r_{\star}(r)}}{\sqrt{f(r)}}.  \label{conformal factor1}
\end{equation}
In addition, the geodesic distance $L(a,b)$ between two points $a$ and $b$ in the conformal flat metric \eqref{conformal flat} is given by
\begin{equation}
L^2(a,b) = \frac{1}{\Omega(a)  \Omega(b)}   \Big( U(b) -U(a) \Big)  \Big( V(a)-V(b) \Big).  \label{distance}
\end{equation}

\section{Page Curves for Non-extremal Kerr-Newman Black Holes} \label{sec3}
\renewcommand{\theequation}{3.\arabic{equation}}\setcounter{equation}{0}

\par In this section, we provide the explicit calculation of the entanglement entropy in Kerr-Newman spacetime. For simplicity, we only consider the quantum field in the non-extremal case \eqref{2d metric1} in this section. In addition, it is essential to emphasize some concrete calculation details. It is widely recognized that deriving the analytical expression  of entanglement entropy in four-dimensional or higher dimensional spacetime poses a challenging task. We typically employ the s-wave approximation \cite{sch} to neglect the contributions from the angular direction, thereby allowing us to concentrate on the dominance of the radial direction. Therefore, the effective two-dimensional theory \eqref{2d metric1} derived from the near-horizon approximation well conforms to this point. Namely, the behavior of the entanglement entropy for Kerr-Neman black holes is described by this theory \eqref{2d metric1} equivalently in the near-horizon region. On the other hand, we assume that the radiation region is described by conformal fields with the central charge $c$. To disregard the back-reaction of Hawking radiation on spacetime, we further assume that black holes are semi-classical, i.e., the relationship between the mass $M$ and the central charge satisfies: $1 \ll c \ll M \sim \frac{1}{G_N}$. Under these assumptions, the dynamics of radiation region are subject to the CFT$_2$. Finally, by neglecting the gray-body factor of Hawking radiation, the entanglement entropy in four-dimensional Kerr-Newman spacetime can be approximately obtained by CFT$_2$.

\subsection{Entanglement Entropy without Island}

\par
\begin{figure}[htb]
\centering
\includegraphics[scale=1.2]{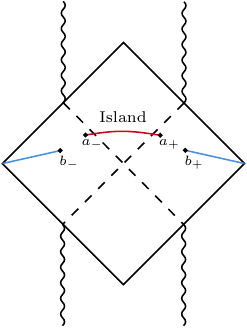}
\caption{\label{kn1} The Penrose diagram for non-extremal Kerr-Newman black holes by considering the entanglement island. The blue line represents the region of radiation where the Hawking radiation is collected by the asymptotical observer. The red line represents the region of island. The points $a_{\pm} /  b_{\pm}$ are denoted by endpoints of island/radiation.}
\end{figure}

 We now assume that black holes formed by the pure state and calculate the entanglement entropy. The Penrose diagram is shown in \mpref{kn1}. We first consider the construction without entanglement island. In this case, only the radiation are left in the whole spacetime. We denote the boundary point for radiation are $b_{\pm}$. The coordinate for $b_+$ is $(\tau, r)=(t_b,r_b)$, and for $b_-$ is $(\tau,r)= (-t_b+\frac{i \beta}{2},r_b)$, where $\beta=\frac{1}{T_H}$ is the inverse temperature \eqref{temperature}.

\par In the absent of island, we need to calculate the interval of radiation $(-\infty,b_-) \cup (b_+, \infty)$. Based on the complementary of entanglement entropy. The entropy in this interval equals to the complementary interval $(b_-,b_+)$. Therefore, the entanglement entropy for a single interval in CFT$_2$ is determined by the logarithmic law \cite{eeformula}
\begin{equation}
\begin{split}
S_{\text{CFT}} (R) &= \frac{c}{3}  \log L(b_-,b_+)  \\
&= \frac{c}{3}  \log  \frac{\Big[U(b_-) - U(b_+) \Big]  \Big[V(b_+) - V(b_-)   \Big]  }{\Omega(b_-)  \Omega(b_+)}.  \label{without island1}
\end{split}
\end{equation}
We substitute the related expression \eqref{conformal factor1} and the coordinate of $(b_-,b_+)$, yield to
\begin{equation}
S_{\text{CFT}} (R) =\frac{c}{6}  \log \frac{4f(r_b)}{\kappa^2} \cosh^2 (\kappa t_b).  \label{without island2}
\end{equation}
Evidently, the expression without island as a function of time $t_b$. At late times, the above expression is approximated as
\begin{equation}
\begin{split}
S_R^{\text{no-island}} &= S_{\text{CFT}}(R)  =\frac{c}{6}  \log  \bigg[ \frac{4(r_b-r_+)(r_b-r_-)(r_+^2 +a^2)^2}{(r_b^2 +a^2)(r_+-r_-)^2} e^{2\kappa t_b}  \bigg] \\
&\simeq \frac{2c}{3} \pi T_H t_b.  \label{result1}
\end{split}
\end{equation}

\par Therefore, the entanglement entropy without entanglement island increases linearly with time $t_b$, which will eventually exceed the entropy bound \cite{bound} and cause information loss. For an eternal black hole, its entanglement entropy is limited to at most twice of the Bekenstein-Hawking entropy. But the result \eqref{result1} demonstrates that entropy without island is infinite at late times. The paradox is sharpened here. In the next subsection, we recalculate the entanglement entropy by taking into account the island. It will result in the expected unitary Page curve for non-extremal Kerr-Newman black holes.

\subsection{Entanglement Entropy with Island}

\par Now, we consider that a single island is contained in Kerr-Newman spacetime. As shown in \mpref{kn1}. We set the coordinates of the island region $(a_-,a_+)$ are $(t_a,r_a)$ for $a_-$ and $(-t_a+\frac{i\beta}{2},r_a)$ for $a_+$. According to the island formula \eqref{island formula}, the contribution of entanglement entropy to the matter part originates from the union $I \cup R$, which is given by \cite{eeformula}
\begin{equation}
S_{\text{CFT}} (R \cup I)= \frac{c}{6}  \log  \bigg[ \frac{L(a_+,a_-) L(b_+,b_-)  L(a_+,b_+)  L(a_-,b_-)}{L(a_+,b_-) L(a_-,b_+)} \bigg].  \label{with island1}
\end{equation}
Then the first area term is written as $\frac{\text{Area} (\partial I)}{4G_N} = 4\pi (r_a^2+a^2)$ \eqref{area}. Correspondingly, the generalized entropy is read off
\begin{equation}
\begin{split}
S_{\text{gen}}     &=  2\frac{\text{Area}(\partial I)}{4G_N}  +S_{\text{CFT}} (R \cup I)  \\
& = \frac{2\pi (r_a^2 +a^2)}{ G_N}  +\frac{c}{6}  \log \bigg[ \frac{16f(r_a)f(r_b)}{\kappa^4}  \cosh^2 (\kappa t_a) \cosh^2(\kappa t_b)  \bigg]  \\
&+ \frac{c}{3}  \log  \bigg[  \frac{\cosh[\kappa(r_{\star}(r_a)-r_{\star}(r_b) )]  -\cosh [\kappa(t_a-t_b)]}{\cosh[\kappa(r_{\star}(r_a)-r_{\star}(r_b) )]  +\cosh [\kappa(t_a+t_b)]}   \bigg]. \label{with island2}
\end{split}
\end{equation}
Here we substitute the expressions \eqref{conformal factor1} and \eqref{distance} into the above equation to simplify. The next step is to extremize the generalized entropy and find the location of island. However, prior to proceeding further, it is essential for us to study the generalized entropy in the early stage.

\par At early times, we assume that the time $t_a$ and $t_b$ are small enough: $t_a \simeq t_b \ll \kappa r_b$. So the entropy \eqref{with island2} behaves as
\begin{equation}
S_{\text{gen}} (\text{early}) \simeq \frac{2\pi (r_a^2 +a^2)}{ G_N}  + \frac{c}{6} \log  \bigg[ \frac{16f(r_a) f(r_b)}{\kappa^4} \cosh^2 (\kappa t_a) \cosh^2(\kappa t_b)   \bigg].  \label{early entropy}
\end{equation}
By extremizing the above expression with respect to the time $t_a$ and the position $r_a$, we consequently obtain
\begin{subequations}
\begin{align}
\frac{\partial}{\partial t_a}  S_{\text{gen}} (\text{early}) &= \frac{c\kappa}{3} \tanh (\kappa t_a)=0,  \\
\frac{\partial}{\partial r_a}  S_{\text{gen}} (\text{early})  &= \frac{4\pi r_a}{G_N}  + \frac{c}{6}  \frac{a^2 (2r_a-r_+-r_-)+r_a(r_a(r_++r_-)-2r_+r_-)}{(r_a^2+a^2) (r_a-r_+) (r_a-r_-)}=0.
\end{align}
\end{subequations}
At the leading order ${\cal  O} (G_N^{-1})$, we find the coordinates of QES at early times is
\begin{equation}
t_a=0,  \qquad r_a \simeq \frac{cG_N}{24G_N} \frac{r_++r_-}{r_+r_-} = \frac{cM}{12\pi a^2} \ell_p^2,  \label{location1}
\end{equation}
where  $\ell_p= \sqrt{G_N}$ is denoted as the four-dimensional Planck length. Our calculation should exclude the physics at the Planck scale. Therefore, indeed, the island is nonexistent at early times. The entanglement entropy is only determined by the radiation \eqref{result1}. Namely, the construction without island always lead to information paradox.

\par
\begin{figure}[htb]
\centering
\includegraphics[scale=1]{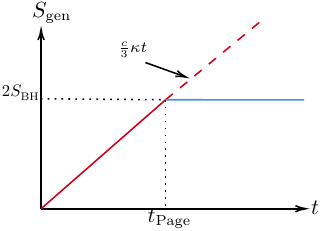}
\caption{\label{pagecurve} The time evolution of entanglement entropy of non-extremal eternal Kerr-Newman black holes. The red line represents the entropy without island. While the blue line represents the entropy with a single island. The Page curve is represented by the solid line.}
\end{figure}

 Next, we turn our attention to the later stage of evaporation. The island that emerges at late times constitute the necessary and sufficient condition for the existence of Page curves. At late times, the time scales $t_a$ and $t_b$ are big enough: $t_a,b \gg \kappa r_b$. Due to the fact that the distance between the left wedge $R_-$ and the right wedge $R_+$ becomes significantly large at this moment. The following approximation can be obtained
\begin{equation}
L(a_+,a_-) \simeq L(b_+,b_-) \simeq L(a_+,b_-) \simeq L(a_-,b_+) \gg L(a_{\pm}, b_{\pm}).  \label{ope}
\end{equation}
Under this approximation, the generalized entropy \eqref{with island2} is reduced to
\begin{equation}
S_{\text{gen}} (\text{late}) = \frac{2\pi (r_a^2 +a^2)}{ G_N} +\frac{c}{6}  \log \bigg[ \frac{4f(r_a)f(r_b)}{\kappa^4} \bigg( \cosh \big( \kappa(r_{\star}(r_a) -r_{\star}(r_b))\big)- \cosh \big(\kappa (t_a-t_b)\big) \bigg)^2   \bigg]. \label{late entropy}
\end{equation}
Similarly, extremizing this expression with respect to time $t_a$ first
\begin{equation}
\frac{\partial}{\partial t_a}  S_{\text{gen}} (\text{late})  = -\frac{c}{3}  \frac{\kappa \sinh[\kappa (t_a-t_b)]}{\cosh[\kappa \big(r_{\star}(r_a) -r_{\star}(r_b) \big) ]  -\cosh[\kappa (t_a-t_b)]} =0.  \label{wrt t}
\end{equation}
By solving this equation, we find that $t_a=t_b$. Invoking the relation $t_a=t_b=t$ into the expression \eqref{late entropy} and extremize it with respect to $r_a$
\begin{equation}
\frac{\partial}{\partial r_a}  S_{\text{gen}} (\text{late}) =\frac{4\pi r_a}{G_N} + \frac{c}{6}  \frac{a^2(2r_a-r_+-r_-)+r_a(r_a(r_++r_-)-2r_+r_-)}{(r_a^2 +a^2)(r_a-r_+)(r_a-r_-)}  -\frac{c\kappa (r_a^2+a^2)}{3(r_a-r_+)(r_a-r_-)}  \Big(1+ \frac{2}{e^{\kappa x}-1} \Big)=0, \label{wrt a}
\end{equation}
where $x \equiv r_{\star}(r_b) - r_{\star}(r_a)$. Subsequently, we take the near-horizon limit: $a \simeq r_+$, yield to
\begin{subequations}
\begin{align}
f(r)      &\simeq f^{\prime} (r_+)(r-r_+) + {\cal O} [(r-r_+)^2] = 2\kappa (r-r_+) + {\cal O} [(r_-r_+)^2]. \label{approxmiation1} \\
r_{\star}(r) &= \int \frac{dr}{f(r)} \simeq \frac{1}{2\kappa} \log \frac{|r-r_+|}{r_+}.   \label{approxmiation2}
\end{align}
\end{subequations}
Substituting these relations into the equation \eqref{wrt a}, we obtain the following equation
\begin{equation}
\frac{\partial}{\partial r_a}  S_{\text{gen}} (\text{late}) \simeq 24 \pi \kappa r_a(r_a-r_+) -2cG_N \kappa e^{-\kappa r_{\star}(r_b)} +\frac{a^2-r_+(r_+-2r_-)}{(a^2+r_+^2)^2}  (r_a-r_+) cG_N=0. \label{location2}
\end{equation}
At last, the location of island is obtained by solving this equation
\begin{equation}
\begin{split}
r_a &\simeq r_+ +\frac{c^2 G_N^2}{ 144 \pi^2  r_+^3}  \frac{r_b-r_-}{r_b-r_-} e^{\frac{-r_b(r_+-r_-)}{r_+^2+a^2}}  +{\cal O}[(cG_N)^3] \\
& \simeq r_+ +\frac{c^2}{144 \pi^2 r_+^3}  \ell_p^2.  \label{location3}
\end{split}
\end{equation}
We find that the distance at which the boundary of island extends beyond the event horizon is limited to the Planck scale, which conforms the near-horizon approximation \eqref{approxmiation1} \eqref{approxmiation2}. Correspondingly, the entanglement entropy of radiation after considering the contribution of island is
\begin{equation}
\begin{split}
S_R^{\text{island}} &\simeq \frac{2\pi (r_+^2 +a^2)}{G_N}  + \frac{c}{6}  \log \bigg[  \frac{4c^2 G_N^2 (r_b-r_+)(r_b-r_-)(a^2+r_+^2)}{9 \pi^2 r_+^3 (r_+-r_-)^3 (a^2+r_b^2)} \bigg]  \\
&= 2S_{\text{BH}} + {\cal O} (cG_N).  \label{result2}
\end{split}
\end{equation}
Namely, the entanglement entropy is dominated by the area term $\frac{\text{Area}(\partial I)}{2G_N}$ at late times, which is consistent with the Bekenstein-Hawking entropy bound. The remaining sub-leading term arise from the contribution of the matter field and can be neglected compared to the leading term. Therefore, entropy with entanglement island eventually asymptotes to twice the Bekenstein-Hawking entropy of an eternal Kerr-Newman black hole. Combing the prior findings without entanglement island \eqref{result1}, we summarize the behavior of entanglement entropy: In the early stage, the entropy increases approximately in a linear manner. In the later stage, the growth of entropy cease. Eventually, the entropy of Hawking radiation is bounded by the Bekenstein-Hawking entropy, which is consistent with the unitarity. Therefore, the information paradox of non-extremal Kerr-Newman black holes can be solve by plotting the corresponding Page curve as shown in \mpref{pagecurve}.

\subsection{Page Time and Scrambling Time}

\par
\begin{figure}[htb]
\centering
\subfigure[\scriptsize{}]{\label{tp1}
\includegraphics[scale=0.35]{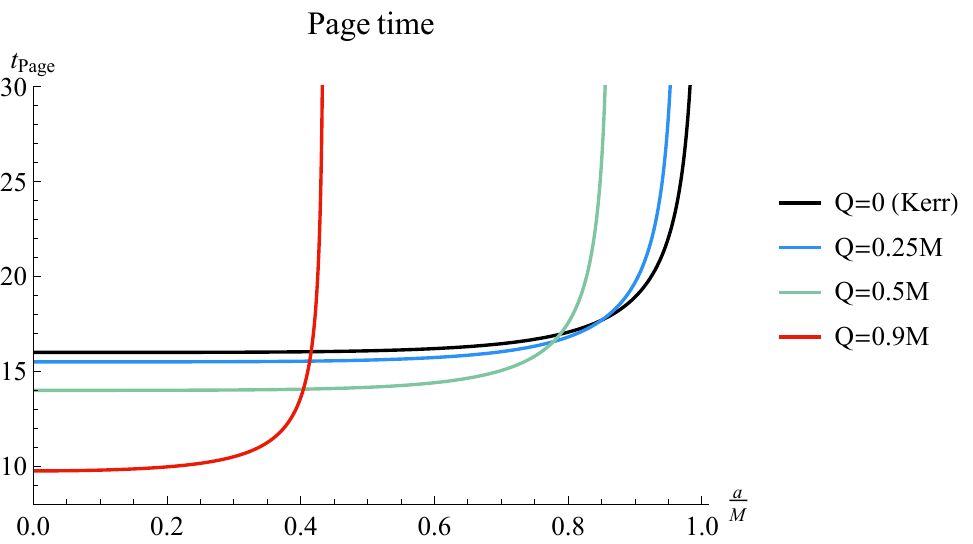}
}
\qquad \qquad \qquad
\subfigure[\scriptsize{}]{\label{tp2}
\includegraphics[scale=0.35]{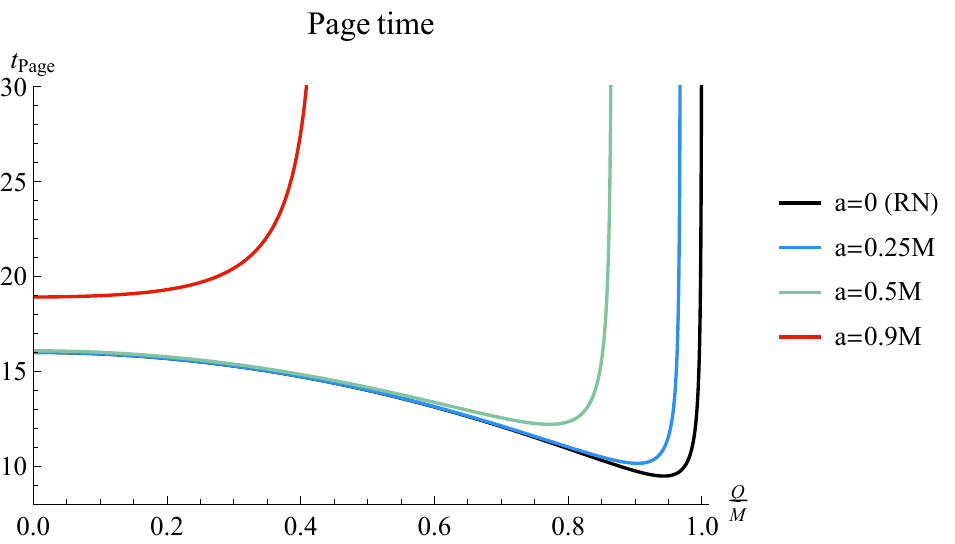}
}
\caption{The Page time as a function for the angular momentum $a$ and the charge $Q$ (in the unit of $\frac{6\pi}{cG_N})$. On the left, the charge $Q$ is fixed; On the right, the angular momentum $a$ is fixed. Note that in the extremal case, the Page time is divergent.}
\label{tp}
\end{figure}

 Finally, we provide the Page time and the scrambling time as by-products of Page curves. The Page time is defined by the moment when the entanglement entropy reduces maximum. For an evaporating black hole, its entanglement entropy will decrease after the Page time. While for an eternal black hole, the entanglement entropy keeps a saturation value after this time. We can determine the Page time by comparing the entropy without island \eqref{result1} and the entropy with island \eqref{result2}:
\begin{equation}
\begin{split}
t_{\text{Page}} (a,Q)  &= \frac{6}{c \kappa} S_{\text{BH}} =\frac{3\beta}{\pi c} S_{\text{BH}}  \\
&=\frac{12\pi}{c G_N}  \frac{(r_+^2+a^2)^2}{(r_+-r_-)}  =\frac{6\pi}{cG_N}  \frac{\Big[ a^2 + \big( M+\sqrt{M^2-a^2-Q^2} \big)^2\Big]^2}{\sqrt{M^2-a^2-Q^2}}  \label{page time}
\end{split}
\end{equation}
In particular, for the special case when $a=0$ (RN black holes) and $Q=0$ (Kerr black holes), our results are consistent with the previous work \cite{rn,kerr}. This further corroborates the validity of our calculations. Then we plot the Page time as a function of the angular momentum $a$ and the charge $Q$ in \mpref{tp}.
We find that the Page time increases as the angular momentum $a$ increases, but decreases as the charge $Q$ increases. However, when the charge or the angular momentum reaches a sufficiently large value, the Kerr-Newman black holes becomes extremal. At this stage, the Page time exhibits divergent. A concrete discussion of the extremal case is provided in the subsequent section.

\par
\begin{figure}[htb]
\centering
\subfigure[\scriptsize{}]{\label{ts1}
\includegraphics[scale=0.3]{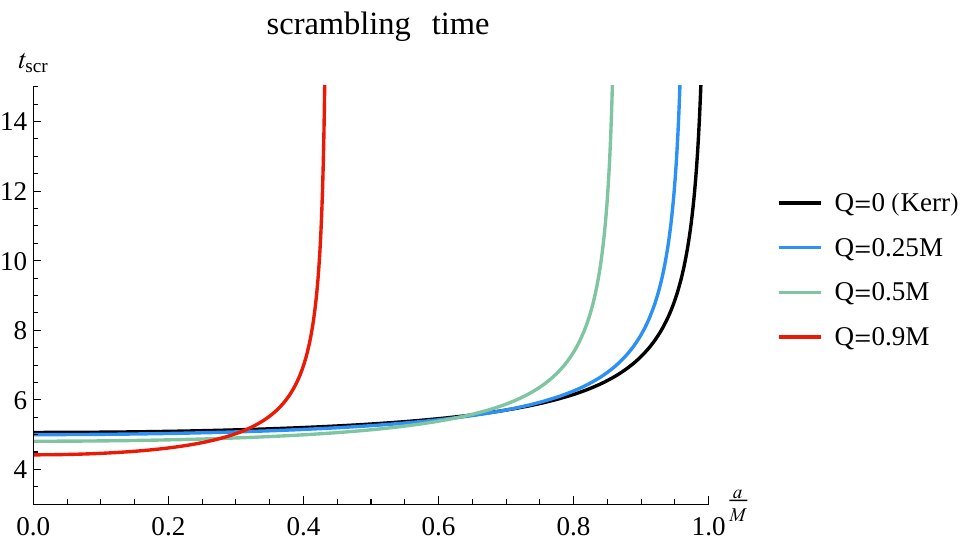}
}
\qquad
\subfigure[\scriptsize{}]{\label{ts2}
\includegraphics[scale=0.3]{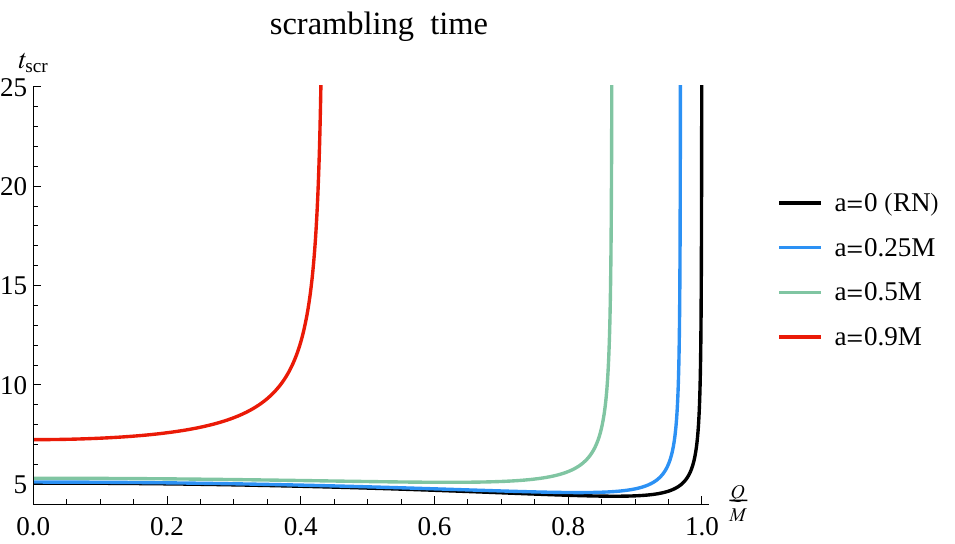}
}
\qquad
\subfigure[\scriptsize{}]{\label{ts3}
\includegraphics[scale=0.3]{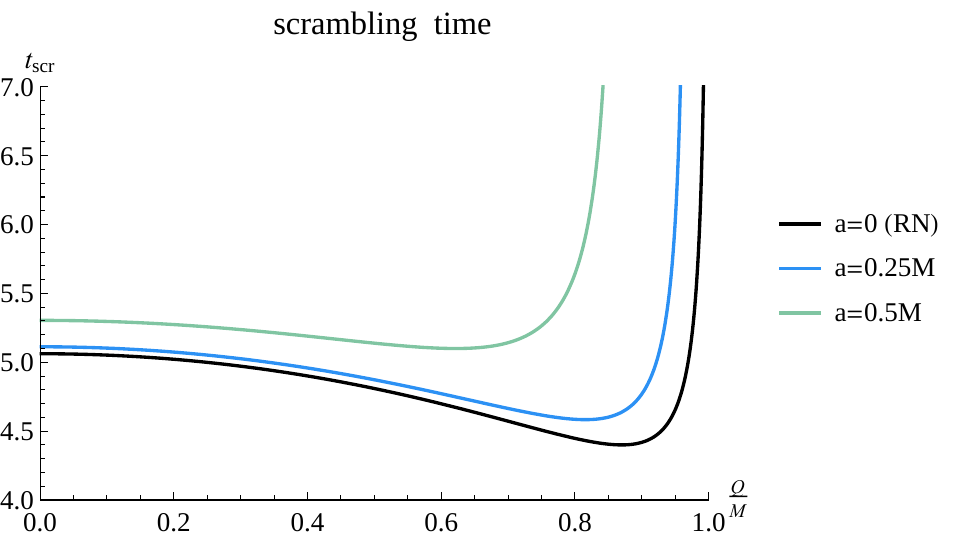}
}
\caption{The scrambling time as a function for the angular momentum $a$ and the charge $Q$. Here we set $G_N=1$. (a) The charge is fixed. (b) The angular momentum is fixed. (c) is the zoomed plot of (b). }
\label{tscr}
\end{figure}

 Further, we discuss the scrambling time. Based on the Hayden-Preskill experiment, for an observer located outside the event horizon, (s)he must wait for the so-called ``scrambling time'' to recover the quantum information that has fallen into the black hole via the emitted Hawking radiation \cite{hp1}. On the other hand, according to the entanglement wedge reconstruction proposal \cite{entanglement wedge}, the scrambling time corresponds to the time when the information reaches the boundary of entanglement islands. Since the entanglement wedge of black hole interior constitutes a portion of the entanglement wedge associated with external radiation. Thus, at time $t_1$, suppose an observer at the cut-off surface $r=r_b$ transmits a light signal toward the black hole. Assuming that the information carried by this signal can be instantaneously decoded upon entering black holes. Then the time at which the information reaches the island $r=r_a$ is denotes as $t_2$. In the null direction, the geodesic distance between these two events is written as
\begin{equation}
v(t_1,r_b) -  v(t_2,r_a) = [t_1 + r_{\star}(r_b)] -[t_2 + r_{\star} (r_a)].  \label{scrambling1}
\end{equation}
 There, the scrambling time is defined by the shortest time interval $\Delta  t= t_2 -t_1$
 \begin{equation}
 \begin{split}
 t_{\text{scr}} \equiv \text{min} [\Delta t] &= \text{min}  \bigg[ \big( r_{\star} (r_b) -r_{\star} (r_a) \big) - \big( v(t_1,r_b) -v(t_2,r_a)\big) \bigg]  \\
 &=r_{\star}(r_b)  -r_{\star} (r_a),  \label{scrambling2}
 \end{split}
 \end{equation}
We take the location of island \eqref{location3} into the expression and obtain
\begin{equation}
 \begin{split}
 t_{\text{scr}} (a,Q) &= \frac{a^2+r_+^2}{2(r_+-r_-)} \log \bigg[ \frac{144\pi^2 (r_b-r_+)(r_+^2+a^2)^2}{c^2 G_N^2 (r_b-r_-)} \bigg]  \\
 &\simeq \frac{1}{2\kappa} \log S_{\text{BH}} + \frac{1}{2\kappa} \log \bigg[\frac{12\pi(r_b-r_+)}{c (r_b-r_-)}  \bigg] \\
 &\simeq  \frac{\Big[ \big( M+ \sqrt{M^2-a^2-Q^2} \big)^2 +a^2 \Big]}{2\sqrt{M^2-a^2-Q^2}}  \log \Bigg[ \frac{\pi \big[a^2+(M+\sqrt{M^2-a^2-Q^2})^2 \big]}{G_N}  \Bigg].   \label{scrambling time}
 \end{split}
 \end{equation}
At the leading order, the result $\frac{1}{2\kappa} \log S_{\text{BH}}$ is agree with the Hayden-Preskill experiment \cite{hp2,hp3}. The scrambling time is logarithmically less the the Page time \eqref{page time}. Therefore, it can be neglected throughout the whole evaporating lifetime. Similarly, for the special cases $a=0$ and $Q=0$, the above results is also consistent with pervious studies \cite{rn,kerr}. We also plot the function of the scrambling time as the function of $Q$ and $a$ in \mpref{tscr}. The behavior of scrambling time exhibits similarities to that of the Page time. It increases as the angular momentum rises and decreases as charges increases. In the same way, the scrambling time also becomes divergent in the extremal case. We leave this point for discussion in the next section.

\section{Near Horizon of Near Extremal Black Holes}  \label{sec4}
\renewcommand{\theequation}{4.\arabic{equation}}\setcounter{equation}{0}

\par Up to now, we have reproduce the Page curve for the non-extremal Kerr-Newman black hole and obtain the Page time \eqref{page time} and the scrambling time \eqref{scrambling time}. However, it is evident from \mpref{tp} and \mpref{tscr} that when the charges or angular momentum is large to render the black hole becomes near extremal, the corresponding physical quantities are divergent and ill-defined. Due to the fact that the distinct spacetime structures of extremal black holes and non-extremal black holes, there exist fundamental differences between the two types \cite{extremalbh3}. Furthermore, For a near extremal black at low or vanishing temperature, a serious problem exists: The internal energy $E$ increases quadratically from the extremal value, namely, $E \sim \frac{T_H^2}{M_{\text{gap}}}$. This suggests the presence of a mass gap of order $M_{\text{gap}}$ in the microscopic spectrum. Consequently, it is expected that a near extremal black hole at temperatures below this gap scale cannot emit the Hawking quantum in the canonical ensemble, where the charges are fixed \cite{massgap}. Therefore, in this section, we focus on the near extremal case and carefully investigate the behavior of entanglement entropy in this scenario. Finally, we provide the precise expressions of entropy, Page time and scrambling time in the near extremal case.

\par
\begin{figure}[htb]
\centering
\includegraphics[scale=1]{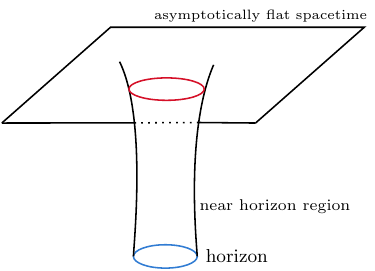}
\caption{\label{warpads} The sketch of the spatial slice of the near extremal Kerr-Newman black hole. At low enough temperature, the near-horizon region is well approximated by the metric \eqref{warp ads}. While far away the horizon, this approximation is invalid and the metric becomes to the asymptotically flat Kerr-Newman geometry.}
\end{figure}

 Firstly, we consider the case of near extremal black holes. Then the corresponding results for extremal cases can be derived by the limiting analysis. Note that the island is actually located in close proximity to the horizon \eqref{location3}. Therefore, we are interested in the near-horizon Kerr-Newman geometry. In this situation, these exists a significant correspondence, which is called the Kerr/CFT correspondence \cite{kerrcft}. This conjecture suggests that under specific boundary conditions, quantum gravity theory in the near-horizon near extremal Kerr geometry is dual to a two-dimensional chiral CFT. This originated from the investigation of asymptotic symmetry groups in the near-horizon geometry of near extremal Kerr black holes. By imposing a definite boundary condition on the asymptotic behavior of the metric, the $U(1)_L$ symmetry of the $SL(2,R)_R \times U(1)_L$ isometry group in the geometry is enhanced to the Virasoro algebra. For further support regarding this field, one can refer to \cite{symmetry1,symmetry2}. In this context, the near-horizon near extremal Kerr black hole closely resemble that of a non-extremal warped AdS$_3$. Correspondingly, due to the presence of the warp AdS$_3$ structure in the near-horizon region of near extremal Kerr-Newmann black holes, the information encoded in its dual CFT can be easily accessed. One can still acquire the entanglement entropy in CFT$_2$ precisely. In addition, the logarithmic law \eqref{without island1} is not applicable when the observer at the near-horizon region. Instead, the entropy follows an area law \cite{eeformula}
\begin{equation}
S_{\text{CFT}} (R \cup I) = - \gamma c \frac{\text{Area}(r)}{L^2(a,b)}, \label{area law}
\end{equation}
where $\gamma$ is a constant.  Back to the gravitational region, we start from the near extremal Kerr-Newman black hole. In order to derive the near horizon geometry for this case, we first consider the following co-rotating coordinate with the angular velocity $\Omega_H$ at the horizon
\begin{equation}
\phi  \to \tilde{\phi} + \frac{a}{r_0^2+a^2} \tilde{t},  \label{corotating coordinate}
\end{equation}
where $r_0=r_{\pm}=M=\sqrt{a^2+Q^2}$ represents the radius of event horizon for the extremal Kerr-Newman black holes.  After here, we use the subscript 0 to represent various parameters in extremal cases. Then we take the near-horizon and near extremal limit with the $\epsilon \to 0$ as follows
\begin{equation}
r \to r_0 +\epsilon \tilde{r},  \quad t \to \frac{r_0^2 + a^2}{\epsilon} \tilde{t},  \quad M_0 \to r_0 + \epsilon^2 \frac{B}{2r_0}.   \label{near extreme}
\end{equation}
Here the parameter $B=\frac{r_+ - r_-}{2 \epsilon}$ is fixed. Finally, the near-horizon geometry can obtained by this scaling limit \cite{symmetry2,warped}
\begin{equation}
d \tilde{s}^2 = \Gamma(\theta) \bigg[ -(\tilde{r}^2-B^2) d\tilde{t}^2 + \frac{d\tilde{r}^2}{(\tilde{r}^2-B^2)} + d\tilde{\theta}^2 \bigg] + \Lambda(\tilde{\theta}) (d \tilde{\phi} +b \tilde{r} d\tilde{t}^2),  \label{warp ads}
\end{equation}
with
\begin{equation}
\Gamma (\theta) = r_0^2 + a^2 \cos^2 \theta, \quad \Lambda(\theta) = \frac{(r_0^2+a^2) \sin^2 \theta}{r_0^2 + a^2 \cos^2 \theta },  \quad  b= \frac{2ar_0^2}{r_0^2 + a^2}.  \label{warp factor}
\end{equation}
Without loss of generality, we set $\tilde{\theta}=0$ to simplify. Then $\Gamma(0)= r_h^2 + a^2$, $\Lambda(0)=0$. The warped AdS$_3$ metric \eqref{warp ads} becomes
\begin{equation}
\begin{split}
d \tilde{s}^2 \bigg|_{\theta=0} &= \Gamma(0) \bigg[ -(\tilde{r}^2-B^2) d\tilde{t}^2 + \frac{ d\tilde{r}^2}{\tilde{r}^2-B^2} \bigg] \\
              &=- (r_0^2 + a^2)(\tilde{r}^2-B^2) d\tilde{t}^2 +  \frac{ (r_0^2 + a^2)d\tilde{r}^2}{\tilde{r}^2-B^2} .  \label{ads3}
\end{split}
\end{equation}
The above spacetime \eqref{ads3} contains a warped AdS$_3$ structure, which allows a dual CFT to describe the near extremal Kerr Newman black at near-horizon region. The structure of spacetime is illustrated in \mpref{warpads}.

\par On the other hand, for near extremal black holes, corrections to their thermodynamics at low temperature constitute another significant study \cite{thermodynamics1}. At the low-temperature limit, the entropy correction of near extremal black holes primarily arises from two sources: At the classical level, this manifests as a linear temperature correction term. An additional quantum correction arises from the zero mode contribution associated with the one-loop determinant. The zero mode here is associated with large diffeomorphisms that preserve the asymptotic structure near AdS$_2$. Up to now, two distinct zero modes have been identified. One linked to the Schwarzian dynamics characterizing the asymptotic region of AdS$_2$, and the other corresponding to fluctuations in the angular velocity of black holes \cite{thermodynamics2}. This relationship can be understood through the dimensional reduction on a deformed two-sphere. The first mode arises from the gravitational fluctuation in AdS$_2$, while the second constitutes the zero mode of the $U(1)$ Maxwell field that originate from the four-dimensional gauge field after the dimensional reduction. These findings have significantly influenced the analysis of semi-classical corrections to the thermodynamics of Kerr-Newman black holes. After considering of these two aspects, the near extremal entropy is obtained in the canonical ensemble at fixed angular momentum as follows \cite{thermodynamics3}
\begin{equation}
\tilde{S}_{\text{BH}} (\tilde{T}_H) = S_0 + a_1 \log S_0 + a_2 \log \frac{\tilde{T}_H}{T_q}, \label{bh entropy2}
\end{equation}
where $S_0 = \frac{\mathcal{A}_0}{4G_N}$is the extremal entropy, the coefficient $a_1$ and $a_2$ is constants of order ${\cal O}(1)$, $\tilde{T}_H = \frac{B}{2\pi}$ is the effective temperature at low temperature limit, and $T_q$  is denoted a emergent scale in the IR comes from the Schwarzian dynamics. At this scale, the excitation energy of the black hole above extremality  is comparable to the average energy of Hawking radiation \cite{thermodynamics4}
\begin{equation}
T_q = \frac{\pi}{ G_N M_0 S_0}. \label{temperature3}
\end{equation}
 It is also noteworthy that as the parameter $B$ approaches $0$, all physical quantities converge to those corresponding to the extremal case. Thus the parameter $B$ serves as a measure of deviation from the extremal case. Thus we will subsequently take the limit of $B \to 0$  to derive the corresponding results in the extremal case.

\par Now we reevaluate the entanglement entropy in spacetime \eqref{ads3} to capture a more comprehensive understanding for near extremal black holes. The Kruskal transformation \eqref{kruskal} is still allowed\footnote{For the near extremal case, note that the surface gravity $\kappa =B=\frac{r_+-r_-}{2\epsilon}$ in the Kruskal coordinate \eqref{kruskal} can not directly derived from the original expression \eqref{surface gravity}. }. After the conformal map, the above metric \eqref{ads3} is converted as follows
\begin{equation}
d \tilde{s}^2 = - \tilde{\Omega}^{-2} (r) dU dV,  \label{conformal flat2}
\end{equation}
with the new conformal factor and the new tortoise coordinate
\begin{equation}
 \tilde{\Omega}^{2}  (r)=  \frac{B^2 e^{2B \tilde{r}_{\star}(\tilde{r})} }{(\tilde{r}^2 -B^2)(r_0^2 +a^2)},  \quad   \tilde{r}_{\star}(\tilde{r})= \frac{1}{2B} \log \frac{\tilde{r}-B}{\tilde{r}+B}.   \label{conformal factor2}
\end{equation}

\par now we reevaluate the entanglement entropy in spacetime \eqref{conformal flat2} to capture a more comprehensive understanding for extremal black holes. We also assume that the near extremal black hole is in a pure state at the initial time $t_0=0$. For the no-island configuration, the entanglement entropy dominated only by the radiation $R$, which is still bounded by $b_{\pm}$. We still need to use \eqref{without island1} to calculate the entropy due to the fact that the QES at this construction ($r_a=0$) is far from the near-horizon region. Then, similar to \eqref{result1}, the entropy is expressed as
\begin{equation}
\begin{split}
\tilde{S}_R^{\text{no-island}} &= \tilde{S}_{\text{CFT}} (R) = \frac{c}{3}  \log  [L(b_+,b_-)] \\
                               &\simeq \frac{2c}{3}  \tilde{T}_H   \tilde{t}_b. \label{result3}
\end{split}
\end{equation}
 So we still maintain the entropy without island increases linearly with time even in the near extremal case. Moreover, we find that in the extremal limit ($B \to 0$), this entropy becomes to zero.

\par  We now consider the island configuration. We anticipate that the entanglement entropy of extremal Kerr-Newman black holes can retain bounded at a finite value by the island formula. After the Page time, an island region $I$ is introduced, and its boundary is also denoted as $a_+$. Fortunately, based on the symmetry of the spacetime, we can investigate the complementary region $(a_+,b_+)$, which contains no singularity \cite{extremalbh1,extremalbh2}. Accordingly, the geodesic distance $L(a_+,b_+)$ is well defined and yields the correct entanglement entropy.

Here we now provide the explicit calculation. Accordingly, the generalized entropy approximates to
\begin{equation}
\begin{split}
\tilde{S}_{\text{gen}}  &= 2\frac{\text{Area}(\partial I)}{4G_N} - 2\gamma c \frac{\text{Area}(\tilde{r}_b)}{L^2(a_+,b_+)} \\
                             &= \frac{2\pi (\tilde{r}_a^2 + a^2)}{G_N} +2a_1 \log(\tilde{r}_a^2 + a^2) + 2a_2 \log \frac{\tilde{T}_H}{T_q}   \\
                             &-2\gamma c \frac{2 \pi \tilde{r}_b^2}{(r_0^2 + a^2)\sqrt{(\tilde{r}_a^2 -B^2)(\tilde{r}_b^2 -B^2)} \Big[ \cosh[\kappa(\tilde{r}_{\star}(\tilde{r}_a) - \tilde{r}_{\star}(\tilde{r}_b)  ) ]  - \cosh [\kappa (\tilde{t}_a - \tilde{t}_b)] \Big]}.  \label{late entropy2}
\end{split}
\end{equation}
We first extremize this expression with respect to time $\tilde{t}_a$, which yields
\begin{equation}
\frac{\partial}{\partial \tilde{t}_a} \tilde{S}_{\text{gen}}   \propto \sinh \bigg( \frac{\tilde{t}_a - \tilde{t}_b}{r_h} \bigg)=0.  \label{wrt t2}
\end{equation}
This suggest that $\tilde{t}_a=\tilde{t}_b$. Then substituting the relation into the expression \eqref{late entropy2} and taking the partial derivative with respect to $\tilde{r}_a$ at the near-horizon limit: $\tilde{r}_a \simeq r_h$, we obtain
\begin{equation}
\begin{split}
\frac{\partial}{\partial \tilde{r}_a}  \tilde{S}_{\text{gen}}   &= \frac{4\pi \tilde{r}_a}{G_N} + \frac{4a_1 \tilde{r}_a}{a^2 +\tilde{r}_a^2}  \\
&- \frac{2c \pi \tilde{r}_a (B-\tilde{r}_b) (B+\tilde{r}_b) \tilde{r}_b^2  \gamma  \text{csch} \left[  \frac{\kappa}{2} \left(  \frac{1}{\tilde{r}_b} -\frac{1}{\tilde{r}_a}  \right)  \right]^2 }{(a^2 + r_0^2)  \tilde{r}_a^3 \tilde{r}_b^3}  =0.  \label{wrt a2}
\end{split}
\end{equation}
By solving this equation in the near-horizon limit, we obtain the location of the island is
\begin{equation}
\tilde{r}_a \simeq r_h + \frac{G_N (c \pi (B-\tilde{r}_b ) \gamma - 2 r_0^3 a_1 )}{2r_0^2 (a^2+r_0^2) \pi}.  \label{location4}
\end{equation}
Finally, the entanglement entropy with island for extremal case is given by
\begin{equation}
\tilde{S}_{R}^{\text{island}} =  \frac{2\pi (\tilde{r}_0^2 + a^2)}{G_N} +2a_1 \log(\tilde{r}_0^2 + a^2)  +{\cal O}(G_N) \simeq 2\tilde{S}_{\text{BH}}.  \label{result4}
\end{equation}
The entanglement entropy in the near extremal case is approximates to the Bekenstein-Hawking entropy at the leading order at late times.  Then, we can determine the Page time for near extremal black hole by \eqref{result3} and \eqref{result4}
\begin{equation}
\begin{split}
\tilde{t}_{\text{Page}} =& \frac{3  \tilde{S}_{\text{BH}}} {c \pi \tilde{T}_H} = \frac{3}{c \pi \tilde{T}_H} \left(S_0 + a_1 \log S_0 +{\cal O}\left( \log \frac{T_q}{\tilde{T}_H}\right) \right).  \\
& \simeq t_0 + t_{\text{Page}}^{\prime} > t_{\text{Page}}.   \label{pagetime2}
\end{split}
\end{equation}
Here $t_0$ is the Page time without considering logarithmic correction \eqref{page time} and $t_{\text{Page}}^{\prime} \sim \frac{\log S_0}{ \tilde{T}_H}$ originates from the logarithmic correction to the extremal entropy \eqref{bh entropy2}.  In addition, the temperature correction term $\log \frac{T_q}{\tilde{T}_H}$ is neglected under the low temperature limit $\tilde{T}_H \sim T_q$. Compared with the previous results \eqref{page time}, it is observed that, in the case of near extremal black holes, the Page time is significantly delayed. This suggests that Schwarzian dynamics predominates in the near-horizon geometry of near extremal black holes and gives rise to quantum corrections to this result. Meanwhile, the scrambling time in near extremal cases can likewise be derived by the similar calculation
\begin{equation}
\begin{split}
\tilde{t}_{\text{scr}}&= \tilde{r}_{\star}(\tilde{r}_b)  -\tilde{r}_{\star}(\tilde{r}_a ) = \frac{1}{2B} \log \frac{(\tilde{r}_b -B)(\tilde{r}_a+B)}{(\tilde{r}_b+B)(\tilde{r}_a -B)} \\
&\simeq \frac{1}{\tilde{r}_a} - \frac{1}{\tilde{r}_b}  \simeq \frac{\tilde{r}_b - r_0}{r_0 \tilde{r}_b} \sim {\cal O}(G_N) > t_{\text{scr}} \sim {\cal O}(\log G_N).
\label{scrambling time2}
\end{split}
\end{equation}
Similar to the Page time at near extremal case \eqref{pagetime2}. The scrambling time at this case is also be delayed compare to the previous result \eqref{scrambling time}. Accordingly, we perform a more rigorous reanalysis of the behavior of Page time and scrambling time by employing near-horizon analysis of near extremal Kerr-Newman black holes, and provide explanations for the dependencies illustrated in \mpref{tp} and \mpref{tscr}. In particular, the entropy is still a finite value \eqref{result4}. Therefore, the island can still yield a finite entropy and ensure the unitary in extremal cases. This also implies the significance and necessity of the island paradigm.

\section{Discussion and Conclusion}  \label{sec5}
\renewcommand{\theequation}{5.\arabic{equation}}\setcounter{equation}{0}

\par In summary, we study the information paradox in the four-dimensional Kerr-Newman spacetime. Due to the fact that the Kerr-Newman black hole represents a non-spherically symmetric higher dimensional spacetime \eqref{kn metric1}, we initially prove that the quantum field in this spacetime can be equivalently described by an effective two-dimensional theory \eqref{2d metric1} in the near-horizon region. Then the entanglement entropy can be well approximated by CFT$_2$ in this framework. According to the island paradigm, the fine-grained entropy of Hawking radiation corresponds to the minimum value among the extremal values of the generalized entropy. We first concentrate on the non-extremal black hole. At early times, black holes have just formed, no island structures are present \eqref{location1}. This leads to the entanglement entropy is contributed by the radiation region and increases linearly with time \eqref{result1}, which sharpen the information paradox. By introducing the entanglement island at late times, the entanglement entropy is eventually dominated by the area term and gradually reaches the saturated Bekenstein-Hawking entropy \eqref{result2}. Based on these findings, we successfully reproduce the Page curve in \mpref{pagecurve} and accurately determine both the Page time \eqref{page time} and the scrambling time \eqref{scrambling time}. We also investigate the impact of the charge $Q$ and the angular momentum $a$ on these physical quantities (\mpref{tp} and \mpref{tscr}). When the charge $Q$ is fixed, both the Page time and the scrambling time increases as the angular momentum increases. Conversely, when the angular momentum $a$ is fixed, both the Page time and the scrambling time decreases as the charge inverses. In particular, for the critical cases when $a=0$ (RN black holes) and $Q=0$ (Kerr black holes), our results are consistent with \cite{rn,kerr}. However, when the black hole reaches to the near extremal case, both the Page time and the scrambling time are approaching divergent. Therefore, we further investigate the situation of near extremal Kerr-Newman black holes. According to the Kerr/CFT correspondence \cite{kerrcft}, the geometry of the near extremal Kerr-Newman black hole in the near-horizon limit is equivalent to a warped AdS$_3$ \eqref{warp ads}. Then we can still correctly evaluate the entropy to adapt the CFT$_2$  in the background for fixed $\theta$ \eqref{ads3}. Furthermore, in the framework of this near extremal near-horizon geometry, the region incorporates an AdS$_2$ component characterized by enhanced $SL(2, R)$ symmetry. This structure accommodates a set of zero modes associated with large diffeomorphisms and gauge transformations, giving rise to strong coupling effects that modify the partition function by one-loop effects \cite{thermodynamics1,thermodynamics2}. So the thermodynamic quantities with logarithmic correction are obtained \eqref{bh entropy2} \eqref{temperature3}.  The we reevaluate the entanglement entropy under these corrections. Similar to the non-extremal case, in the configuration without island, the entanglement entropy still grow linearly with time even in the extremal case \eqref{result3}. Subsequent, by taking account of islands, we conclude that the entanglement entropy is still bounded by the Bekenstein-Hawking entropy by the island formula even in the extremal case with vanishing temperature \eqref{result4}. Significantly, we also find that both the Page time \eqref{pagetime2} and the scrambling time \eqref{scrambling time2} for near extremal black holes are delayed. This suggests that the near extremal limit not only corresponds to the geometric formation of AdS$_2$ in the near-horizon region of the Kerr-Newman black hole, but also signifies the domain dominated by Schwarzian dynamics, reflecting the quantum fluctuation of the gravitational system. These conclusions aligns with the principle that the entanglement entropy of a finite system remains finite, which also implies the validity of the island formula. Our study broadens the application scope of the island formula and offers a systematic calculation method of the Page curve in the most general stationary spacetime. This work holds great potential value for future studies.

\par For the future research, a nice motivation is to study the evolution of entanglement entropy of evaporating Kerr-Newman black holes. It is anticipated that the emergence of islands will lead to the entropy drops to zero at the end of evaporation. Another interesting aspect to explore is the comtribution of superradiance to the Page curve. The superradiance of BTZ black holes has been studied previously in \cite{ymhbtz}. However, the context of our study involves asymptotically flat spacetime, which eliminates the necessity for a coupled thermal bath to absorb the Hawking radiation. This feature renders our framework more realistic and  representative. We intend to consider the superradiance into our analysis to further study the effects of charges and angular momentum on the Page curve. Such study will enhance our understanding of the information paradox and the island paradigm.

\begin{acknowledgments}
We would like to thank Jun Nian, Yu Tian and Yi Ling for helpful discussions. The study was partially supported by NSFC, China (Grant No.12275166 and No.12311540141).
\end{acknowledgments}

\end{document}